\documentclass[reprint, floatfix,amsmath, aps, pra, physics, nofootinbib]{revtex4-1}

\usepackage{graphicx}
\usepackage{bm}
\usepackage{physics}
\usepackage[dvipsnames]{xcolor}
\usepackage[linkcolor=cyan]{hyperref}
\usepackage{svg}
\usepackage{amsmath}
\usepackage{amsfonts}
\usepackage{float}

\newcommand{\hilb}{\mathcal{H}}

\begin{document}

\title{Parameterized Hamiltonian simulation using quantum optimal control}

\author{Paul Kairys}
\email{pkairys@vols.utk.edu}
\affiliation{Bredesen Center for Interdisciplinary Research and Graduate Education, University of Tennessee, Knoxville, Tennessee}
\affiliation{Quantum Science Center, Oak Ridge National Laboratory, Oak Ridge, Tennessee}
\thanks{This manuscript has been authored by UT-Battelle, LLC, under contract DE-AC05-00OR22725 with the US Department of Energy (DOE). The US government retains and the publisher, by accepting the article for publication, acknowledges that the US government retains a nonexclusive, paid-up, irrevocable, worldwide license to publish or reproduce the published form of this manuscript, or allow others to do so, for US government purposes. DOE will provide public access to these results of federally sponsored research in accordance with the DOE Public Access Plan (http://energy.gov/downloads/doe-public-access-plan).}

\author{Travis S.~Humble}
\affiliation{Bredesen Center for Interdisciplinary Research and Graduate Education, University of Tennessee, Knoxville, Tennessee}
\affiliation{Quantum Science Center, Oak Ridge National Laboratory, Oak Ridge, Tennessee}

\begin{abstract}
Analog quantum simulation offers a hardware-specific approach to studying quantum dynamics, but mapping a model Hamiltonian onto the  available device parameters requires matching the hardware dynamics. We introduce a paradigm for quantum Hamiltonian simulation that leverages digital decomposition techniques and optimal control to perform analog simulation. We validate this approach by constructing the optimal analog controls for a superconducting transmon device to emulate the dynamics of an extended Bose-Hubbard model. We demonstrate the role of control time, digital error, and pulse complexity, and we explore the accuracy and robustness of these controls. We conclude by discussing the opportunity for implementing this paradigm in near-term quantum devices.
\end{abstract}

\maketitle

\section{Introduction}

Estimating dynamical properties of entangled many-body quantum states is critical for scientific discovery and engineering robust quantum technologies \cite{georgescu_quantum_2014,alexeev2021quantum}. Unfortunately, even the most sophisticated algorithms are limited to simulating quantum dynamics at small times due to the exponential growth in memory requirements with increasing quantum state entanglement \cite{schuch2008entropy,lacki_dynamical_2019,rizzi_simulation_2008}. Quantum computers, on the other hand, are capable of efficient simulation of quantum dynamics because even highly entangled quantum states can be represented efficiently in a quantum register \cite{nielsen2002quantum}.
\par
Simulating the dynamics of a model quantum system using a controllable quantum device is referred to as quantum simulation \cite{georgescu_quantum_2014}. Two leading paradigms for quantum simulation are based on digital and analog representations. In digital quantum simulation, sequences of discrete quantum logic gates are compiled to generate unitary operators that implement time evolution of the model system, while in analog quantum simulation the quantum device is controlled so that its own evolution imitates the dynamics of the model system \cite{georgescu_quantum_2014}.
\par
Several approaches have combined digital quantum logic with analog quantum evolution, especially for quantum simulation \cite{salathe_digital_2015,parra-rodriguez_digital-analog_2020,lamata_digital-analog_2018,martin_digital-analog_2020,celeri_digital-analog_2021,lamata_digital-analog_2017,arrazola_digital-analog_2016,galicia_enhanced_2020,babukhin_hybrid_2020}. These approaches augment digital simulation methods with specialized analog evolution to extend the types of quantum states that can be prepared. However, a typical hurdle for analog quantum simulation is the design or discovery of a controllable quantum device whose Hamiltonian is isomorphic to the model Hamiltonian \cite{georgescu_quantum_2014}. In many cases, construction of the device itself limits this isomorphism to small regime of model parameters and, therefore, the class of possible simulations.
\par
Here we address this limitation by introducing a framework to decompose analog quantum simulation into a sequence of individually controlled unitaries. By designing sequences of unitaries with optimal control, this framework demonstrates as a direct route to generating model dynamics of interest. This generalizes recent work by Holland et al., which used optimal control for analog simulation of nuclear physics \cite{holland_optimal_2020}. In addition, we show that the resulting decomposition partitions the problem of identifying optimal analog controls into a linear set of independent optimization problems. This enhances the computational efficiency of designing the simulations and expands the class simulatable models. In justifying the feasibility of the  optimal control in a simulation paradigm, we introduce a strategy to realize quantum simulation over a wide range of model parameters without additional control optimizations. 
\par 
We structure the remainder of the work as follows: in Sec.~\ref{sec:tqs} we describe our proposed simulation paradigm. In Sec.~\ref{sec:demo} we demonstrate the feasibility of the paradigm in three steps: Sec.~\ref{subsec:application} defines the model system of interest, Sec.~\ref{subsec:device_ham} defines the controllable device, and Sec.~\ref{subsec:results} discusses the numerical results of our demonstration. Finally we conclude and discuss our contributions in Sec.~\ref{sec:discussion}.

\section{Parameterized simulation with optimal control}\label{sec:tqs}

Quantum simulations require specification of a model quantum system and a controllable quantum system. Typically this is accomplished by defining a parameterized model Hamiltonian $H_M(\vec{\lambda})$ and a parameterized, time-dependent device Hamiltonian $H_D(\vec{\alpha},t)$. The parameters of the device Hamiltonian $\vec{\alpha}$ represent controllable parameters such as tunable electric or magnetic fields. The parameters of the model Hamiltonian $\vec{\lambda}$ are specified over some finite range $\lambda_{l,min} \leq \lambda_l \leq \lambda_{l,max}$ on which interesting dynamical physics are sought.
\par
The task of quantum simulation is that the dynamics of the quantum device over a control time $T_c$,
\begin{align}
    U_D(\vec{\alpha},T_{c}) = \mathcal{T}\exp \bigg[-\frac{i}{\hbar}\int_0^{T_c} H_D(\vec{\alpha},\tau) d\tau \bigg],
\end{align}
and the dynamics of the model system over a time $T_s$,

\begin{align}
    U_M(\vec{\lambda},T_s) = \exp \bigg[-\frac{iT_s}{\hbar}H_M(\vec{\lambda}) \bigg],
\end{align}
are equivalent. In practice, this equivalence is replaced by requiring the dynamics to be sufficiently close according to some distance measure. We consider a measure between two unitary operators given by the infidelity
\begin{equation}\label{eq:infidelity}
    g(U_1,U_2) = 1 - \frac{1}{\text{dim($U$)}} \bigg\vert \textrm{Tr}(U^\dagger_1 U_2) \bigg\vert ^2 
\end{equation}
which is invariant to a global phase and takes a positive value between zero and one \cite{palao_optimal_2003}.
\par
We assume the Hamiltonians are defined on a composite Hilbert space of $N$ $d$-dimensional quantum systems, $\mathcal{H} = \bigotimes_i^N \mathcal{H}_i$, where each Hamiltonian is a sum of $k$-local operators. A $k$-local operator acts as the identity on all but $k$ of the $d$-dimensional local Hilbert spaces, $\mathcal{H}_i$. We will assume that the model Hamiltonian has the form
\begin{equation}\label{eq:target_hamiltonian}
    H_M(\vec{\lambda}) = \sum_{l=0}^L \lambda_l H_{M,l},
\end{equation}
where $H_{M,l}$ is $k$-local. Then, we consider a device Hamiltonian composed as a linear combination of operators with the same locality as the model Hamiltonian,
\begin{align}\label{eq:device_ham}
    H_D(\vec{\alpha},t) = \sum_{l=0}^L H_{D,l}(\vec{\alpha},t),
\end{align}
i.e., both $H_{M,l}$ and $H_{D,l}$ act on the same $k$-local subspace $l$. Furthermore, we assume that the subsystems acted on by each local term are addressable via a chosen set of device parameters, i.e., $\forall~ l ~\exists~ \vec{\alpha}$ s.t. $H_{D,l'}(\vec{\alpha},t)=0~\forall~ l'\neq l$. This structure of the device Hamiltonian ensures that any subspace on which model evolution occurs can be addressed independently by a choice of device parameters. 
\par 
One approach to realizing quantum simulation is to numerically determine the parameters $\vec{\alpha}$ for a control time $T_c$ that generates the target unitary dynamics for a specific parameter set $\vec{\lambda}$ and model simulation time $T_s$ as
\begin{align}
    \min_{\vec{\alpha}} g(U_D(\vec{\alpha},T_c),U_M(\vec{\lambda},T_s)).
\end{align}
There are two consequence of this approach. First, the model and device unitaries become exponentially large with increasing quantum system size. Second, for every desired set of parameters $\vec{\lambda},T_s$, a new set of control parameters and control time must be determined. This strategy is more than exponentially inefficient. 
\par
We overcome these bottlenecks by assuming the model Hamiltonian parameters are bounded and discretely parameterized. Explicitly, we assume that each parameter is bounded as $\lambda_{l,min} \leq \lambda_l \leq \lambda_{l,max}~\forall~l$ and discretized as a grid with step size $\Delta \lambda_l >0$ for each parameter $l$ such that $\lambda_{l} = \lambda_{l,min} + n_l \Delta\lambda_l$. Next, we use the Trotter decomposition to convert the global model unitary operator into a product of $k$-local unitary operators \cite{lloyd1996universal,georgescu_quantum_2014}
\begin{widetext}
\begin{align}\label{eq:tailored_quantum_sim}
    U_M(\vec{\lambda},T_s) &= \exp \bigg[ -\frac{iT_s}{\hbar} H_{M}(\vec{\lambda})\bigg] =\exp \bigg[ -\frac{iT_s}{\hbar} \sum_{l=0}^L \lambda_l H_{M,l}\bigg] =\exp \bigg[ -\frac{iT_s}{\hbar} \sum_{l=0}^L \bigg(\lambda_{l,min} + n_l \Delta \lambda_l\bigg) H_{M,l}\bigg]\\
    \nonumber &= \lim_{q \rightarrow \infty} \bigg[ \prod_{l=0}^L  \exp\bigg(-\frac{iT_s}{q \hbar}  (\lambda_{l,min} + n_l \Delta \lambda_l) H_{M,l} \bigg)\bigg]^q \\
    \nonumber &= \lim_{q \rightarrow \infty} \bigg[ \prod_{l=0}^L  \exp\bigg(-\frac{iT_s}{q \hbar}  \lambda_{l,min} H_{M,l} \bigg)\exp\bigg(-\frac{iT_s}{q \hbar}  \Delta \lambda_l H_{M,l} \bigg)^{n_l}\bigg]^q \\
    & = \lim_{q \rightarrow \infty} \bigg[ \prod_{l=0}^L  U_{M,l}\bigg(\lambda_{l,min},\frac{T_s}{q} \bigg)  U_{M,l}\bigg(\Delta \lambda_l, \frac{T_s}{q} \bigg)^{n_l} \bigg]^q \label{eq:tqs_composable_general}
\end{align}
\end{widetext}
One typically truncates the series at an order $q>>1$ that gives a desirable global error defined by the conservative error bound
\begin{align}
    U_{M}(\vec{\lambda},T_s)- U_{M}^{(q)}(\vec{\lambda},T_s) &= \frac{T_s^2}{2q} \sum_{l>m=1}^{M} [\lambda_l H_l,\lambda_m H_m] \\
    \nonumber &~~~+ \mathcal{O}\bigg(\frac{T_s^3}{q^2}\bigg).
\end{align}
\par
Based on Eq.~(\ref{eq:tqs_composable_general}), a target unitary generated by a Hamiltonian of $L$ parameters requires determining only $2L$ unitaries. Furthermore, to implement these unitary operators in a quantum device, one solves $2L$ optimization problems corresponding to $\lambda_{l,min}$ as
\begin{align}
    \min_{\vec{\alpha}} g(U_{D,l}(\vec{\alpha},T_c),U_{M,l}(\lambda_{l,min},T_s/q)),
\end{align}
and $\Delta \lambda_l$ as
\begin{align}
    \min_{\vec{\alpha}} g(U_{D,l}(\vec{\alpha},T_c),U_{M,l}(\Delta \lambda_{l},T_s/q))
\end{align}
where $U_{D,l}$ is the unitary operator generated by addressing the subspace on which $H_{M,l}$ acts as

\begin{align}
    U_{D,l}(\vec{\alpha},T_{c}) = \mathcal{T}\exp \bigg[-\frac{i}{\hbar}\int_0^{T_c} H_{D,l}(\vec{\alpha},\tau) d\tau \bigg].
\end{align}

\par
The above result demonstrates that using optimal control within the context of quantum simulation is efficient, in principle. In practice, additional structure to the model Hamiltonian permits more efficient implementations. For example, if the model Hamiltonian is sufficiently uniform, $H_{M,l} \approx H_{M,l'} ~\forall~ l,l'$, and the device Hamiltonian is sufficiently uniform, $H_{D,l} \approx H_{D,l'} ~\forall~l.l'$, then one can develop control ansatze that are capable of determining the controls across any subset of the device, requiring only minor parameter refinements (or calibrations) for each subsystem. 
\par
The former condition implies that the model Hamiltonian has internal symmetry as is common in condensed matter or chemical physics \cite{georgescu_quantum_2014}. The latter condition, that the quantum device is relatively uniform, will typically be satisfied from an engineering perspective as it decreases device complexity and cost  \cite{arute_quantum_2019,brown_co-designing_2016}. We discuss extensions and other applications of this approach in Appendix~\ref{app:extensions}.

\section{Demonstration}\label{sec:demo}
We next demonstrate the validity and feasibility of optimal control based quantum simulations through a demonstration: emulating the dynamics of the extended Bose-Hubbard (EBH) model in a device of interacting transmon superconducting circuit elements. 

\subsection{Application to Bose-Hubbard dynamics}\label{subsec:application}

The Bose-Hubbard (BH) model describes spinless bosons moving in a discrete space, like a lattice or graph. The BH model is particularly relevant to studying ensembles of cold atoms in optical lattices, light-matter interactions, multi-particle quantum walks, and dynamical quantum phase transitions \cite{morsch_dynamics_2006,eckardt_colloquium_2017,yan_strongly_2019,lacki_dynamical_2019}. The EBH model includes an additional term that has been shown to be a building block for novel phases of matter including topologically entangled quantum spin liquid phases, making it of immediate interest for near-term quantum simulation \cite{isakov2011topological}. 
\par 
Recently, analog quantum simulations of the BH model have been performed in systems of capacitively-coupled, tunable-frequency transmons \cite{yan_strongly_2019,ye_propagation_2019}. This is possible because capacitively coupled transmons in the dispersive regime have a Hamiltonian isomorphic to the BH Hamiltonian. However, parameter ranges for the simulation are limited due to hardware constraints and the device hardware used in these simulations do not support analog implementations of the EBH model, which is the sum of the typical BH model Hamiltonian with an additional extension term \cite{yanay_two-dimensional_2020}. 
\par
For a one-dimensional chain of lattice sites, the instance of the EBH Hamiltonian considered in this work is
\begin{align}\label{eq:EBH_ham}
    H_{EBH} &= J \sum_{i=0}^{N} (\hat{b}^\dagger_i \hat{b}_{i+1}+ \hat{b}_i \hat{b}^\dagger_{i+1}) + V \sum_{i=0}^{N} \hat{n}_i \hat{n}_{i+1} \\
    &= H_{BH} + H_{E},
\end{align}
where $V$ quantifies the strength of the two-site potential, $J$ quantifies the kinetic energy of particle hopping, and the operators $\hat{b}^\dagger_i, \hat{b}_i,\hat{n}_i$ are the bosonic creation, annihilation operators, and number operators acting on site $i$, respectively. In this representation, $H_{EBH}$, is the sum of two-local operators and thus is a variant of Eq.~(\ref{eq:device_ham}). 
\par
In our demonstration, we explore dynamical properties in the subspace spanned by states of $\{\ket{0}_i,\ket{1}_i\}~\forall ~i$ for various parameter values $V/J = -1,-2,-3$. This is accomplished by fixing $J=-0.1 / 2\pi$ GHz, $V_{min} = \Delta V = 0.1/ 2\pi$ GHz. Then, the Hamiltonian Eq.~(\ref{eq:EBH_ham}) is parameterized by a single, discrete parameter $n_v=1,2,3$ that corresponds the weight $V/J$, according to the decomposition presented in Sec.~\ref{sec:tqs}. 
\par
Following with the approach outlined in Sec.~\ref{sec:tqs}, we transform the global time-evolution operator into a product of two-site unitary operators using a Trotter decomposition as
\begin{align}
    \nonumber U_{EBH}(n_v,T_s) &= \exp\bigg[-\frac{iT_s}{\hbar} (H_{BH}+n_v H_{E}) \bigg] \\
    &= \lim_{q \rightarrow \infty} \bigg[ \prod_{i=0}^{N} U_{BH}^{i,i+1}\bigg(\frac{T_s}{q}\bigg) U_{E}^{i,i+1}\bigg(\frac{T_s}{q}\bigg)^{n_v}\bigg]^q \label{eq:EBH_trott},
\end{align}
where the parameter $n_v$ determines the magnitude of $V/J$, as in Eq.~(\ref{eq:tailored_quantum_sim}). We note that for each pair of sites, $i,i+1$, there is the BH evolution unitary
\begin{equation}\label{eq:BH_trotter_step}
    U_{BH}^{i,i+1}\bigg(\frac{T_s}{q}\bigg) = \exp \bigg[-\frac{i T_s}{\hbar q} J(\hat{b}^\dagger_i \hat{b}_{i+1}+ \hat{b}_i \hat{b}^\dagger_{i+1})\bigg]
\end{equation}
and the EBH potential evolution unitary
\begin{equation}\label{eq:EBH_trotter_step}
    U_{E}^{i,i+1}\bigg(\frac{T_s}{q}\bigg) = \exp \bigg[-\frac{i T_s}{\hbar q} V\hat{n}_i \hat{n}_{i+1}\bigg].
\end{equation}
\par 
As mentioned in Sec.~\ref{sec:tqs} these operators are locally the same for any two sites $i,j$ and $i',j'$. Provided that the device Hamiltonian is sufficiently uniform, we will assume that a suitable control ansatz used to find optimal controls for a single pair of sites will also be valid for any other pair of sites across the whole device. Essentially reducing the optimal control task to the solution of two optimization problems. One is for BH evolution,
\begin{align}\label{eq:min_BH}
    \min_{\vec{\alpha}} g(U_{D,l}(\vec{\alpha},T_c),U_{BH}\bigg(\frac{T_s}{q}\bigg))
\end{align}
and the other one is for EBH potential evolution,
\begin{align}\label{eq:min_E}
    \min_{\vec{\alpha}} g(U_{D,l}(\vec{\alpha},T_c),U_{E}\bigg(\frac{T_s}{q}\bigg)),
\end{align}
where the superscripts labeling sites $i,i+1$ have been removed to emphasize these local unitaries are invariant with respect to site labeling.
\par
The lone hyper-parameter introduced in Sec.~\ref{sec:tqs} is the order of the Trotter decomposition, $q$. An accurate simulation requires that the ratio $\frac{T_s}{q} << 1$. However, varying $q$ also changes the unitaries to determine via optimal control, and we evaluate the role of $q$ on the fidelity of the optimal controls by fixing $T_s=1$ ns and varying the Trotter order $1<q<10$ in Sec.~\ref{subsec:results}.

\subsection{Device Hamiltonian}\label{subsec:device_ham}

\begin{figure*}
    \includegraphics[width=1.0\textwidth]{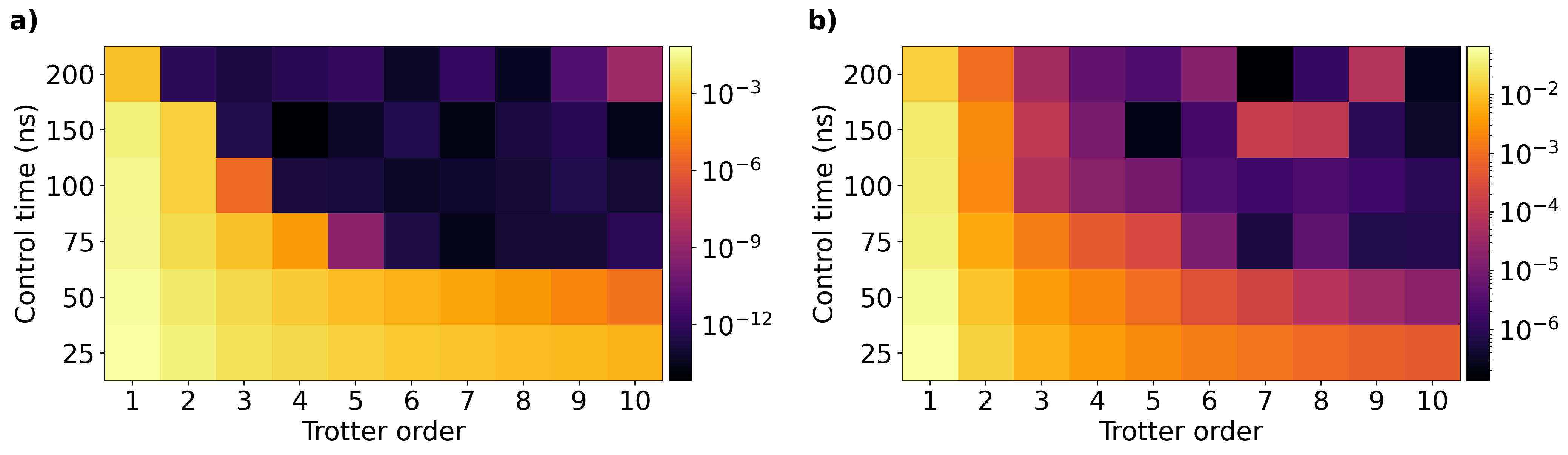}
    \caption{Amplitude constrained infidelities. a, heat-map showing the minimum observed infidelities after optimization for control times (y-axis) and Trotter decomposition order q (x-axis). For each parameter pair (pixel coordinate) 10 optimizations were run with independent initial parameter choices for a control ansatz of 20 Gaussian functions. A q-dependent infidelity is observed demonstrating approximate control times of 75 ns required to obtain arbitrarily low infidelity. b, heat-map showing the average observed infidelities of all 10 optimization instances for each parameter pair. The slow, monotonic decrease in infidelity below $T_c =75$ ns is due to increasing q decreasing the norm of Trotter-step unitary and therefore the norm of the infidelity.}
    \label{fig:amp_constrained_QSL}
\end{figure*}

\begin{figure*}
    \includegraphics[width=1.0\textwidth]{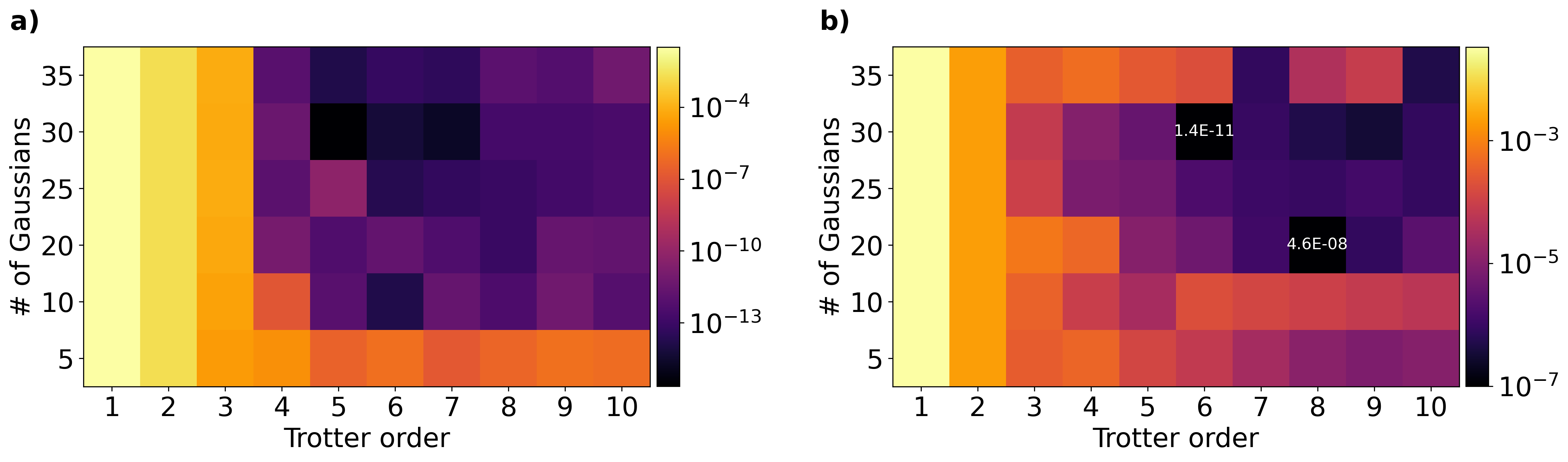}
    \caption{Parameterization constrained infidelities. a, heat-map showing the minimum observed infidelities after optimization for control times (y-axis) and Trotter decomposition order q (x-axis). For each parameter pair (pixel coordinate) 10 optimizations were run with independent initial parameter choices for a control time of 100 ns. We observe that the infidelity has a strong dependence on pulse complexity, as determined by the number of Gaussian basis functions. b, heat-map showing the average observed infidelities of all 10 optimization instances for each parameter pair. The slow, monotonic decrease in the mean infidelity is due to increasing q decreasing the norm of Trotter-step unitary and therefore the norm of the infidelity.}
    \label{fig:param_induced_QSL}
\end{figure*}

\begin{figure*}
    \includegraphics[width=1.0\textwidth]{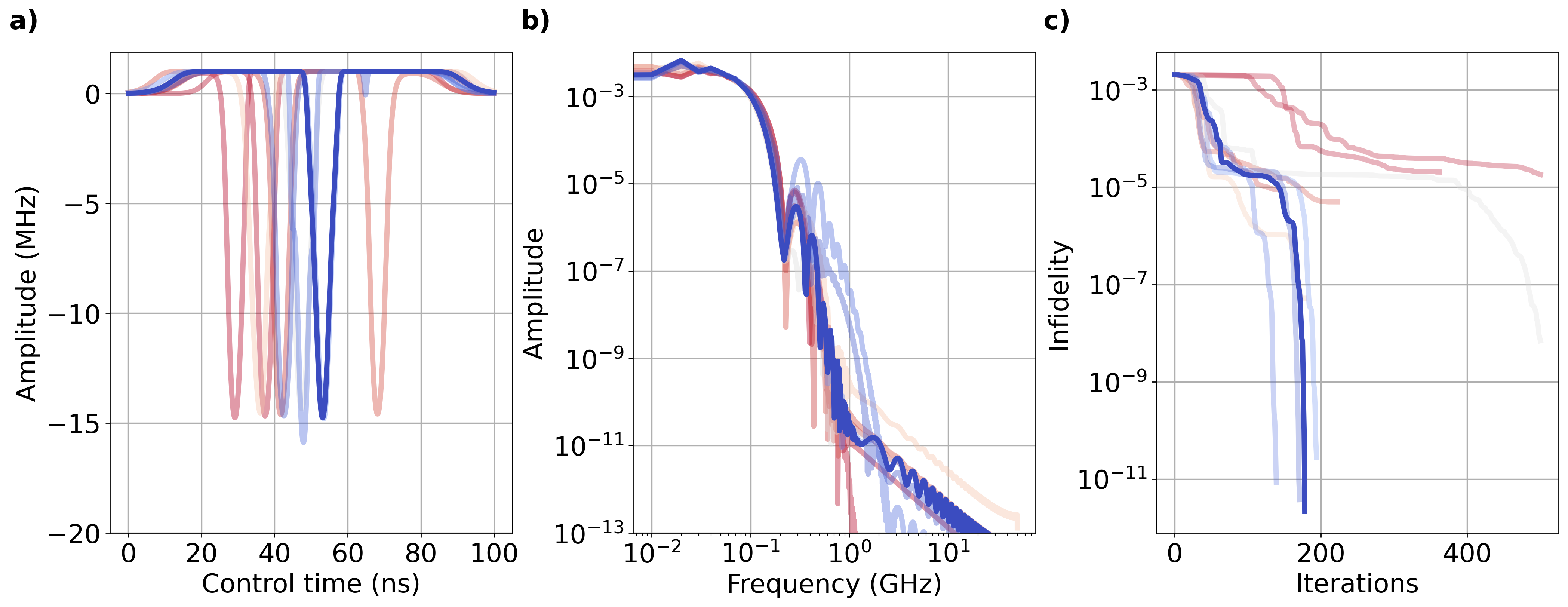}
    \caption{Optimal pulse family. a, family of optimal pulses obtained from different initial guesses. Each optimal pulse used a control time of $T_c=100$ ns and a Trotter decomposition order $q=6$. The GOAT ansatz parameterized the coupling control as a function of 20 Gaussian pulses with variable amplitude, mean, and standard deviation. During the optimization the control amplitude was constrained within the bounds shown in Table~\ref{tab:params}. We observe a saturation of the upper bound and a sharp dip approaching the lower bound as a general topological feature of the pulse family. The control pulses are smooth and start and stops at zero with slow ramp times. b, periodograms for the family of pulses shown in a. For the optimum pulse (darkest blue) spectral density is concentrated within the range of $[0,1]$ GHz. This is expected due to the sharp dip in negative amplitude of the control pulse centered around $55$ ns and the resonant transmon frequencies. c, The convergence of the GOAT optimization routine for various initial guesses. The most optimal pulses converge within 200 iterations however some take longer to converge. The maximum permitted iterations were 500, which was saturated by two optimizations out of the ten run. Detailed information about the optimization procedure can be found in Appendix~\ref{app:GOAT}.}
    \label{fig:opt_pulses}
\end{figure*}

\begin{figure*}
    \includegraphics[width=1.0\textwidth]{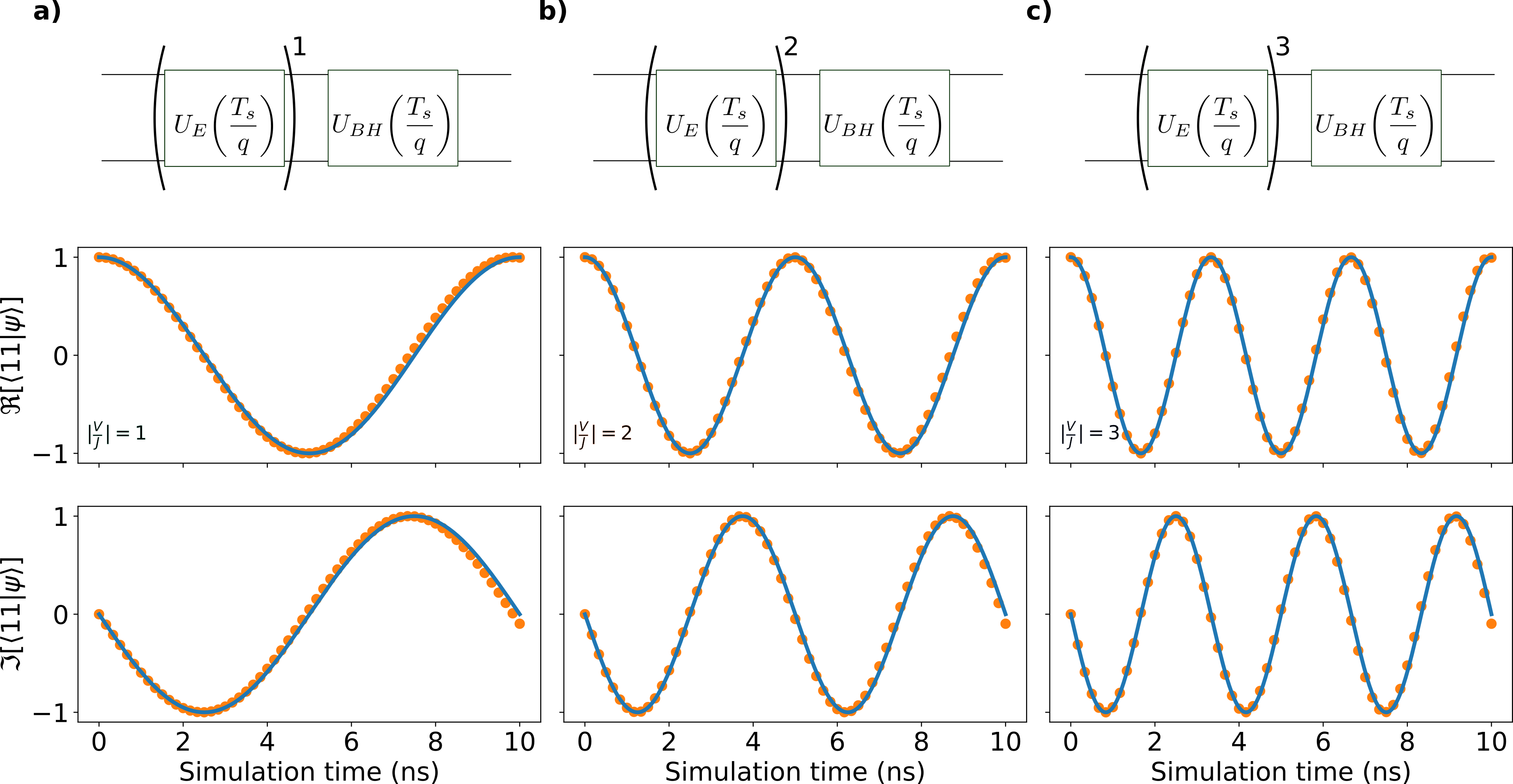}
    \caption{Parameterized Phase tracking. In each column the real (middle row) and imaginary (bottom row) components of the phase accumulated on the $\ket{11}$ state when time-evolved under the EBH Hamiltonian. The Trotterized BH evolution $U_{BH}(1/6)$ is generated using a 50 ns Gaussian pulse, with parameters $a=-2.07$ MHz, $\mu = 25$ ns, $\sigma=6.36$ ns, which we found provided infidelity below $10^{-6}$. The EBH Trotter-step unitary and the BH Trotter-step unitary are then repeatedly applied to the initial state up to $N=60$ times because the Trotterized time step is 1/6 (ns). In each column the number of applications of $U_{E}(1/6)$ is varied so that the equivalent ratio of $V/J=1,2,3$ for columns 1,2, and 3, respectively. This generates the discrete evolution plotted as orange circles and is compared with the exact evolution shown in blue lines. This demonstrates that by selectively compiling the bare pulses that generate the individual Trotter steps, it is possible to reconstruct the dynamics of an EBH Hamiltonian with arbitrary $V/J$.}
    \label{fig:phase tracking}
\end{figure*}

Quantum devices composed of superconducting circuits are a leading paradigm for quantum computing and systems of  tunable-coupling, tunable-frequency transmons are a particularly mature instance of that technology at large scales \cite{arute_quantum_2019,li_tunable_2020,yan_tunable_2018,reagor_demonstration_2018,sung_realization_2020}. Tunable couplers can be used to effectively turn-off interactions between transmons when those interactions are undesired. This allows addressing subsystems of the quantum device with high fidelity; one of the primary assumptions in deriving our main result of Sec.~\ref{sec:tqs} \cite{yan_tunable_2018}.
\par
It is common to model such devices as systems of interacting Duffing oscillators on a planar lattice with a tunable coupling \cite{yan_strongly_2019}. The local Hilbert space of each transmon is a Fock space, isomorphic to the local Hilbert space of a lattice site in the Bose-Hubbard model. Thus we map the model Hilbert space to the device Hilbert space by mapping each lattice site $i$ in the model Hamiltonian Eq.~(\ref{eq:EBH_ham}) to a single transmon $i$. This allows us to write down a device Hamiltonian with the same geometry as the model Hamiltonian as, i.e. a one dimensional chain of $N$ transmons
\begin{align}\label{eq:eff_hamiltonian}
    H_D(\vec{\gamma}) &= \sum_{i=0}^N \omega_{i} \hat{n}_i + \frac{\delta_i}{2} \hat{n}_i(\hat{n}_i - 1) \\
    &+ \omega_{i+1} \hat{n}_{i+1} + \frac{\delta_{i+1}}{2} \hat{n}_{i+1}(\hat{n}_{i+1} - 1) \\
    &+ \gamma_{i,i+1}(t)(a^\dagger_i a_{i+1} + a_i a^\dagger_{i+1}),
\end{align}
where $\omega_i,\delta_i$ are the frequency and anharmonicity of transmon $i$, respectively, and $\gamma_{i,i+1}$ is the controllable (time-dependent) coupling of transmon $i$ to transmon $i+1$. The tunable coupling addresses pairs of transmons individually from the rest of the device. When a pair of transmons interact, the strength of the interaction is related to the the strength of the coupling $\gamma_{i,i+1}$ and the relative detuning of the transmon frequencies $\omega_i-\omega_{i+1}$. The strongest interaction is present when the transmons are tuned to the same frequency, $\omega_{int.}$, which is the case we will consider here \cite{krantz_quantum_2019}. Thus, the Hamiltonian governing dynamics of an interacting two-transmon subsystem is

\begin{align}
    H_D^{i,i+1}(\gamma_{i,i+1}) &= \omega_{int.} \hat{n}_i + \frac{\delta_i}{2} \hat{n}_i(\hat{n}_i - 1) \\
    &+ \omega_{int.} \hat{n}_{i+1} + \frac{\delta_{i+1}}{2} \hat{n}_{i+1}(\hat{n}_{i+1} - 1) \\
    &+ \gamma_{i,i+1}(t)(a^\dagger_i a_{i+1} + a_i a^\dagger_{i+1}).
\end{align}

\setlength{\tabcolsep}{35pt}
\begin{table}
    \centering
    \caption{Parameters used in this work to specify the device Hamiltonian, taken from Ref.~\cite{sung_realization_2020}}
    \begin{tabular}{lc }
        \hline \hline Parameter & Value (GHz)  \\ \hline
        $\omega_{1}/2\pi$  &  4.16 \\
        $\omega_{2}/2\pi$ &  4.00 \\
        $\delta_1/2\pi$  &  -0.220 \\
        $\delta_2/2\pi$ &  -0.210 \\
        $\omega_{int.}/2\pi$  &   4.16 \\
        $\min(\gamma_{12})/2\pi$  &   -0.04 \\
        $\max(\gamma_{12})/2\pi$  &  0.002 \\
        \hline \hline
    \end{tabular}
    \label{tab:params}
\end{table}

Moreover, quantum information processing in transmon systems is performed in reference to a pre-calibrated rotating frame associated with the idling frequency of each transmon. This transformation is given by the unitary transform
\begin{align}
    R_{i,i+1}(t) &= R_i(t) \otimes R_{i+1}(t)\\
    &= \exp \bigg[ \frac{i t}{\hbar} \omega_{idle,i} \hat{n}_i \bigg] \otimes \exp \bigg[ \frac{i t}{\hbar} \omega_{idle,i+1} \hat{n}_{i+1}\bigg]
\end{align}
which yields an effective Hamiltonian in this rotating frame, $H^R_{i,i+1}(\gamma_{i,i+1}) = i\hbar (\partial_t R_{i,i+1}(t)) R_{i,i+1}^\dagger(t) + R_{i,i+1}(t)H^{i,i+1}_D(\gamma_{i,i+1})R_{i,i+1}^\dagger(t)$, and the corresponding time-ordered evolution operator for control time $T_c$ of a two-transmon subsystem is
\begin{equation}\label{eq:device_unitary}
    U_{i,i+1}^R(\gamma_{i,i+1},T_c) = \mathcal{T}\exp \bigg[ -\frac{i}{\hbar} \int_0^{T_c} d\tau H^{R}_{i,i+1}(t,\gamma_{i,i+1}) \bigg].
\end{equation}
\par
As shown in Sec.~\ref{subsec:application}, we decompose the quantum simulation into two minimization problems for each model device subsystem, Eq.~(\ref{eq:min_BH}) and Eq.~(\ref{eq:min_E}). Given the device Hamiltonian for a two-transmon system, we specify these optimization problems as
\begin{align}\label{eq:min_BH_easy}
    \min_{\gamma(t)} g(U^R(\gamma(t),T_c),U_{BH}\bigg(\frac{T_s}{q}\bigg))
\end{align}
and
\begin{align}\label{eq:min_E_hard}
    \min_{\gamma(t)} g(U^R(\gamma(t),T_c),U_{E}\bigg(\frac{T_s}{q}\bigg)),
\end{align}
where we drop site labels for clarity.
\par
We remark that the first optimization problem Eq.~(\ref{eq:min_BH_easy}) is trivial to solve in the computational subspace because the model and device Hamiltonians in this subspace are isomorphic. We use a single Gaussian function describing $\gamma(t)$ to achieve infidelities below $10^{-6}$. 
\par
The non-trivial optimization problem given by Eq.~(\ref{eq:min_E_hard}) requires numerical exploration. We note that Eq.~(\ref{eq:min_E_hard}) is currently framed as a functional minimization problem, of $\gamma(t)$. However, to minimize this function numerically requires expressing $\gamma(t)$ with a finite number of parameters. We choose to do this by decomposing the coupling as the sum of multiple Gaussian functions
\begin{align}
    \gamma(\vec{\alpha},t) = \sum_{m=0}^M a_m \exp\bigg[ \frac{(t-\mu_{m})^2}{2\sigma_{m}^2} \bigg]
\end{align}
where we define the parameter vector $\vec{\alpha} = [a_0,\mu_0,\sigma_0,\dots,a_M,\mu_M,\sigma_M]$. This yields a numerical optimization task
\begin{align}
    \min_{\vec{\alpha}} g(U^R(\gamma(\vec{\alpha},t),T_c),U_{E}\bigg(\frac{T_s}{q}\bigg)),
\end{align}
The total number of Gaussian functions $M$ is varied to understand the impact of control complexity on the optimal infidelities. The optimization methods, including discussion of leakage levels, algorithms, and implementations are detailed in Appendix~\ref{app:methods}.

\subsection{Numerical Results}\label{subsec:results}

A quantum simulation enabled by optimal control requires understanding the relationship between the control time $T_c$ and the Trotter decomposition order $q$. We evaluate this parameter space over a grid of control times $25~\text{ns}\leq T_c \leq 200~\text{ns}$ and Trotter order $1\leq q \leq 10$ by performing an ensemble of optimizations at each $(T_c,q)$ instance. Figure~\ref{fig:amp_constrained_QSL} plots for each pair of parameters (a) the minimum observed infidelity and (b) the mean observed infidelity obtained from each ensemble. In Fig.~\ref{fig:amp_constrained_QSL}a, two distinct regimes appear that correspond to infidelities of approximately $10^{-3}$ and $10^{-12}$. We refer to these regions as as high-infidelity and low-infidelity regimes, respectively. 
\par 
In Fig.~\ref{fig:amp_constrained_QSL}a, we observe that control times below 75 ns are unable to achieve infidelities of $ 10^{-12}$ regardless of the Trotter-decomposition order. This may be related to a quantum speed limit for the controls, which represents a lower bound on the time required to implement a unitary operator with zero infidelity \cite{deffner_quantum_2017,lee_dependence_2018,kirchhoff_optimized_2018}. In fact, a crucial feature of quantum speed limits is observed here: the sharp decrease of the infidelity between regimes as the control time is increased \cite{kirchhoff_optimized_2018}. We also observe that for $T_c \gtrsim 75$ ns, the boundary between high and low-infidelity regimes is a function of the Trotter order $q$, such that larger Trotter orders yield lower infidelities for shorter control times. Hence, the optimal analog quantum simulation of the EBH model on this transmon device is both fast and accurate. 
\par
In Fig.~\ref{fig:amp_constrained_QSL}b, we present the mean of the infidelity. Regions with small minimum infidelities typically have small mean infidelities. We also observe a general decrease in infidelity with increasing $q$. This is due to the Trotter-step unitary, Eq.~(\ref{eq:EBH_trotter_step}), converging to the identity operator with large $q$. The identity operator is a trivial solution for the optimal control problem, thus the controls which generate the identity are a type of attractor for the optimization routine, leading to lower fidelity solutions with increasing $q$.
\par
Our simulations have used control pulses as  linear combinations of Gaussian functions since Gaussian pulses permit control over the pulse bandwidth and are commonly used in experimental superconducting systems \cite{motzoi_simple_2009,mckay_efficient_2017}. However, pulse complexity impacts how accurately the target unitaries are prepared \cite{lloyd_information_2014,lucarelli_quantum_2018}. We evaluate the dependence of infidelity on pulse complexity in Fig.~\ref{fig:param_induced_QSL} by fixing $T_c=100$ ns and varying the number of Gaussian functions from 5 to 35. We did not observe significant changes in the minimum and mean infidelities when using a control pulse with more than 35 Gaussians. 
\par
In Fig.~\ref{fig:param_induced_QSL}a, we show a high-infidelity regime and a low-infidelity regime with a sharp transition from $\approx 10^{-4}$ infidelity to $\approx 10^{-12}$ infidelity, respectively. The boundary between the regime is a function of control complexity and Trotter order. But, the dependence on these parameters is qualitatively different from Fig.~\ref{fig:amp_constrained_QSL} in that given sufficient pulse complexity, the transition between high and low-infidelity regions is only dependent on $q$, and dependence on the number of Gaussians vanishes. This is to be expected, as the optimal configuration of $N$ Gaussians exist as an optimal configuration to a parameterization of $N+1$ Gaussians. 
\par
In Fig.~\ref{fig:param_induced_QSL}b, we observe that increasing the number of Gaussians leads to a decrease in the mean observed infidelity, regardless of Trotter order $q$. However, we observe a discrepancy in the trend in infidelity with increasing $q$ for $N=10$ and $N=35$ Gaussians. These pulse complexities typically have higher mean infidelities with increasing $q$ compared to other pulse complexities $N$, which demonstrates that the optimization landscape is a function of pulse complexity \cite{lee_dependence_2018}.
\par
We present a family of optimal pulses in Fig.~\ref{fig:opt_pulses} that are obtained by running 10 optimizations at control time $T_c=100$ ns and Trotter decomposition order $q=6$ with a control ansatz of 20 Gaussian functions. In Fig.~\ref{fig:opt_pulses}a, we present the time-domain representation of the optimal pulses. We note that the family of optimal pulses share a number of common features: ramps to positive coupling, sharp dips into negative coupling, then back to positive coupling and ramps to zero coupling. In particular, the family of lowest infidelity pulses vary slightly in the location of the negative amplitude peak, indicating some insensitivity of the infidelity on peak location within the range $40-60$ ns. 
\par
Periodograms of the optimal pulse family are shown in Fig.~\ref{fig:opt_pulses}b. Nearly all of the spectral intensity is concentrated in frequency components $\leq1$ GHz. Finally, in Fig.~\ref{fig:opt_pulses}c, we show the convergence of the pulse optimization for each pulse in the family. We observe that the optimal pulses with the best fidelity converged most quickly, typically within 200 iterations.
\par
A final demonstration of the ability to compile optimized analog pulses into sequences permitting simulation with variable Hamiltonian parameters is shown in Eq.~(\ref{eq:tqs_composable_general}). For the EBH model from Eq.~(\ref{eq:EBH_trott}), the unitary evolution governed by an EBH Hamiltonian for different ratios of $|V/J|$ is generated by repeated application of multiple Trotter steps of Eq.~(\ref{eq:EBH_trotter_step}). We use the lowest infidelity pulse from Fig.~\ref{fig:opt_pulses}a that generates evolution of $U_E(T_s/q)$ and compile it with the single Gaussian pulse that generates $U_{BH}(T_s/q)$  for variable $|V/J|$. In Fig.~\ref{fig:phase tracking}, we show the real and imaginary components of the probability amplitude of the state $\ket{11}$ as a function of the number of applied unitaries, $U_E(T_s/q)$. We observe that by compiling pulses found via optimal control, dynamics can be generated for Hamiltonians with variable ratios of $|V/J|$. This evolution is visualized as a quantum circuit for clarity, but represents a smooth analytic pulse formed by concatenating the individual pulses that generate $U_E(T_s/q)$ and $U_{BH}(T_s/q)$.

\section{Discussion and Conclusion}\label{sec:discussion}

Analog and digital quantum simulation are both promising approaches to studying Hamiltonian dynamics of quantum states. While several recent approaches to combine these ideas are driven by digital quantum computation \cite{parra-rodriguez_digital-analog_2020,lamata_digital-analog_2018,holland_optimal_2020}, we have introduced a paradigm that uses optimal quantum control to implement Hamiltonian simulation protocols. First, we have shown how a parameterized Trotter decomposition of a target unitary propagator limits the number of local optimal control problems to a linear function of the device Hamiltonian parameters. We then compiled the resulting optimal pulses to simulate the target Hamiltonian for arbitrary model parameter values.
\par
We have demonstrated Hamiltonian simulation by considering a dynamics simulations of the extended Bose-Hubbard model within a system of superconducting transmons. Notably, the model Hamiltonian is not isomorphic with the target Hamiltonian and the model parameter values lie outside of the device parameter regimes. Using an ensemble of numerical simulations we have explored the role control time, Trotter decomposition order, and control complexity on the obtainable infidelities using these optimal control pulses. 
\par 
Our numerical results show that fast, high-fidelity control pulses are present over a range of Trotter orders and pulse complexities. The discovered optimal controls are smooth and concentrated in the low frequency domain, which supports the future realization in near-term quantum hardware. Together, these results emphasize that compilation of the optimal controls enables simulation of Hamiltonians with arbitrary parameter values. 

\section*{Acknowledgements}
This material is based upon work supported by the U.S.~Department of Energy, Office of Science, National Quantum Information Science Research Centers, Quantum Science Center and the U.S.~Department of Energy, Office of Science, Early Career Research Project. This research used resources of the Compute and Data Environment for Science (CADES) at the Oak Ridge National Laboratory, which is supported by the Office of Science of the U.S. Department of Energy under Contract No. DE-AC05-00OR22725.

\appendix

\section{Extensions}\label{app:extensions}

In Section~\ref{sec:tqs}, we decompose the global unitary time evolution of a model Hamiltonian into a sequence of local unitary evaluations using the Trotter decomposition, and demonstrate how one can further decompose the evolution into a small number of local unitaries when the Hamiltonian is appropriately parameterized. This decomposition is not unique and depending on the application, alternative decompositions may prove to be more practical. Here we consider another example of a parameterized model Hamiltonian in which each the parameters of $\vec{\lambda}$ lie in a finite range around zero: $\lambda_{l,min} \leq 0\leq \lambda_{l,max}$ and the parameter range is discretized on a fixed grid with spacing $\Delta \lambda_l > 0$. Then we can define an arbitrary point in the parameter space by two parameters $0 \leq n_{l,-} \leq N_{l,-}$ and $0 \leq n_{l,+} \leq N_{l,+}$ with an expression $\lambda_l = (n_{l,+} - n_{l,-}) \Delta \lambda_l$. 
\par
Using the above parameterization, we consider the Trotter decomposition of the global unitary operator into a product of $k$-local unitary operators to produce a modified version of Eq.~(\ref{eq:tailored_quantum_sim}) and Eq.~(\ref{eq:tqs_composable_general})

\begin{widetext}
\begin{align}
    U_M(\vec{\lambda},T_s) &= \exp \bigg[ -\frac{iT_s}{\hbar} H_{M}(\vec{\lambda})\bigg] =\exp \bigg[ -\frac{iT_s}{\hbar} \sum_l \lambda_l H_{M,l}\bigg] =\exp \bigg[ -\frac{iT_s}{\hbar} \sum_l \bigg((n_{l,+} - n_{l,-}) \Delta \lambda_l\bigg) H_{M,l}\bigg]\\
    \nonumber &= \lim_{q \rightarrow \infty} \bigg[ \prod_{l=0}^L  \exp\bigg(-\frac{iT_s}{q \hbar}  ((n_{l,+} - n_{l,-}) \Delta \lambda_l) H_{M,l} \bigg)\bigg]^q \\
    \nonumber &= \lim_{q \rightarrow \infty} \bigg[ \prod_{l=0}^L  \exp\bigg(-\frac{iT_s}{q \hbar}  (-\Delta \lambda_l) H_{M,l} \bigg)^{n_{l,-}}\exp\bigg(-\frac{iT_s}{q \hbar}  \Delta \lambda_l H_{M,l} \bigg)^{n_{l,+}}\bigg]^q \\
    & = \lim_{q \rightarrow \infty} \bigg[ \prod_{l=0}^L  U_{M,l}\bigg(-\Delta \lambda_l,\frac{T_s}{q} \bigg)^{n_{l,-}} U_{M,l}\bigg(\Delta \lambda_l, \frac{T_s}{q} \bigg)^{n_{l,+}} \bigg]^q
\end{align}
\end{widetext}

Comparing this decomposition to Eq.~(\ref{eq:tailored_quantum_sim}) and Eq.~(\ref{eq:tqs_composable_general}) we see that depending on the parameterization of the Hamiltonian, a decomposition of the time evolution operator can be chosen to limit the total number of unitary applications per Trotter step. For example, if the goal is to simulate a Hamiltonian with $\lambda_l = 2\Delta \lambda_l$ then the above decomposition requires only two unitaries to be applied per Trotter step, whereas if one were to use Eq.~(\ref{eq:tqs_composable_general}) this same simulation would require $N_{l,-}+2$ applied unitaries per Trotter step.
\par
In addition to selecting a preferred decomposition based on the Hamiltonian's parameterization we also remark on the ability to add local disorder to the quantum dynamics simulation. In this case we can consider that a particular instance of the model Hamiltonian $H_M(\vec{\lambda_l})$ is perturbed by some vector $\vec{\epsilon_l}$. For simplicity, we will assume that each local perturbation $\epsilon_l$ is chosen from a range of perturbation strengths $\epsilon_{l,max} \leq 0 \leq \epsilon_{l,min}$ that are discretized on a grid with spacing $\Delta \epsilon_l >0$. Then any perturbation can be described as $\epsilon_l = (n_{l,\epsilon,+} - n_{l,\epsilon,-}) \Delta \epsilon_l$, using the same notation as introduced above. 
\par
Using the above decomposition we consider the Trotter decomposition of the global unitary operator into a product of $k$-local unitary operators to produce a modified version of Eq.~(\ref{eq:tailored_quantum_sim}) and Eq.~(\ref{eq:tqs_composable_general}):

\begin{widetext}
\begin{align}
    U_M(\vec{\lambda}+\vec{\epsilon},T_s) &= \exp \bigg[ -\frac{iT_s}{\hbar} H_{M}(\vec{\lambda}+\vec{\epsilon_l})\bigg]\\
    \nonumber 
    &= \exp\bigg[- \frac{iT_s}{\hbar} \sum_{l=0}^L \bigg(\lambda_{l,min} + n_l \Delta \lambda_l + (n_{l,\epsilon,+} - n_{l,\epsilon,-}) \Delta \epsilon_l \bigg) H_{M,l}\bigg]\\
    \nonumber &= \lim_{q \rightarrow \infty} \bigg[ \prod_{l=0}^L  \exp\bigg(-\frac{iT_s}{q \hbar}  (\lambda_{l,min} + n_l \Delta \lambda_l+ (n_{l,\epsilon,+} - n_{l,\epsilon,-}) \Delta \epsilon_l ) H_{M,l} \bigg)\bigg]^q \\
    & = \lim_{q \rightarrow \infty} \bigg[ \prod_{l=0}^L  U_{M,l}\bigg(\lambda_{l,min},\frac{T_s}{q} \bigg)  U_{M,l}\bigg(\Delta \lambda_l, \frac{T_s}{q} \bigg)^{n_l}U_{M,l}\bigg(- \Delta \epsilon_{l}, \frac{T_s}{q} \bigg)^{n_{l,\epsilon,-}} U_{M,l}\bigg(\Delta \epsilon_l, \frac{T_s}{q} \bigg)^{n_{l,\epsilon,-}} \bigg]^q \label{eq:disordered_tqs}
\end{align}
\end{widetext}

This result indicates that by adding additional unitaries into each Trotter step of Eq.~(\ref{eq:tqs_composable_general}) yields a quantum simulation of a Hamiltonian that has local disorder, weighted by the number of times each perturbation unitary is applied in each trotter step. For additional clarity we consider a case where $\Delta \lambda_{l} = \Delta \epsilon_l$. Then, in this case one can simplify Eq.~(\ref{eq:disordered_tqs}) into
\begin{align}
    U_M(\vec{\lambda},T_s) &= \lim_{q \rightarrow \infty} \bigg[ \prod_{l=0}^L  U_{M,l}\bigg(\lambda_{l,min},\frac{T_s}{q} \bigg)\cdot \\
    &~~~~~~~~~~~~~U_{M,l}\bigg(\Delta \lambda_l, \frac{T_s}{q} \bigg)^{n_l+n_{l,\epsilon,+} - n_{l,\epsilon,-}} \bigg]^q .
\end{align}
which still requires finding optimal controls for $2L$ unitary operators but permits quantum simulation of Hamiltonians with local disorder
\par
Finally, we have focused on the case of time-independent problem Hamiltonians but in principle they may be time dependent as well. We speculate that in the case of a time-dependent Hamiltonian one could combine our strategy with digital methods for time-dependent Hamiltonians such as the one proposed in Ref.~\cite{poulin2011quantum}, but we do not consider that generalization in this work. 

\section{Methods}\label{app:methods}

\subsection{Problem formulation}

As discussed in Section~\ref{sec:demo} in order to implement a quantum simulation requires determining a set of control parameters $\vec{\alpha}$ such that the unitary operator generated by the device evolution, Eq.~(\ref{eq:device_unitary}), is sufficiently close to a chosen Trotter-step unitary, according to some distance measure. In this work we consider a measure between two unitary operators given by the infidelity:

\begin{equation}
    g(\vec{\alpha}) = 1 - \frac{1}{\text{dim(U)}}  \bigg\vert Tr(U^\dagger_{M} P_{c}U_D(\vec{\alpha},T_c)P_{c}) \bigg\vert ^2
\end{equation}

where $\vec{\alpha}$ is the vector of parameters to optimize over, $P_{c}$ is a projector onto a desired computational subspace, $U_{M}$ is the desired unitary operator to implement in the computational subspace, and $U_D(\vec{\alpha},T_c)$ is the time evolution operator over the full system Hilbert space at time $T_c$ given by Eq.~(\ref{eq:device_unitary}).
\par
In this work we demonstrate quantum simulation by considering a system of two transmons with a tunable interaction. The local Hilbert space of each transmon $\hilb_i$ is an infinite-dimensional Fock space, and the composite space of two transmons $i=1,j=2$, is given by $\hilb_1 \otimes \hilb_2$. However, to perform a computation of these systems requires first choosing a finite subspace of this composite Hilbert space.
\par
In the dispersive regime, the effective Hamiltonian between two transmons interacting via a tunable coupling transmon, is given by Eq.~(\ref{eq:device_ham}). We note that this Hamiltonian is block-diagonal in the total excitation number basis, due to the fact that the interaction term conserves the total number of excitations. Thus, if we are interested in preparing a unitary operator in the qubit subspace spanned by the orthonormal basis $\{\ket{00},\ket{01},\ket{10},\ket{11}\}$, we need only consider the Hilbert space composed of the first three levels of each transmon because this includes the subspace with total excitations $N=0,1,2$ and the device Hamiltonian does not permit population transfer between blocks of different excitation number.

By taking into account the first three levels of each transmon in our simulations any state overlapping $\ket{11}$ will populate the states $\ket{02},\ket{20}$ during the control time. This is due to the fact that these quantum states share a total number of excitations $N=2$. This phenomena is referred to as ``leakage," and is a common issue in superconducting systems. We note that the formulation of the infidelity function, Eq.~(\ref{eq:infidelity}), will penalize unitary operators that, couple $\ket{11}$ to $\ket{02}$ or $\ket{20}$ at the final time $T_c$. However, it will not penalize unitaries that couple these states at intermediate control times $T<T_c$, which is not generally possible given the form of Eq.~(\ref{eq:eff_hamiltonian}). 

Based on these facts we remark that the optimal controls found in this work minimize leakage outside of the computational subspace only at the final time, not intermediate times. At first glance this seems like an unwanted effect. However, due to the device Hamiltonian, dynamics in the $N=2$ excitation subspace are necessary to generate the correct final-time unitary. I.e, the additional basis states $\ket{02},\ket{20}$ are actually being used for information processing in a beneficial way, and are necessary to implement the desired evolution in the computational subspace.

\subsection{Optimal Control with GOAT}\label{app:GOAT}

There are numerous quantum optimal control algorithms for optimizing a problem like the one above, here we focus on the recently developed gradient optimization of analytic controls (GOAT) algorithm \cite{machnes_tunable_2018}. This method was chosen because of it's rapid convergence to optimal solutions and the ability to determine low parameter-count controls which may be easier to calibrate on real quantum devices \cite{machnes_tunable_2018,kirchhoff_optimized_2018}.

The GOAT algorithm considers Hamiltonians that are a linear combination of drift Hamiltonians that are constant and control channels that are parameterized and time-dependent: 

\begin{align}
    H(\vec{\alpha},t) = H_{drift} + \sum_k^K \varepsilon_k(\alpha_k,t)H_k
\end{align}

and expands each driving field in a superposition of functions

\begin{align}\label{eq:control_amp}
    \varepsilon_k(\alpha_k,t) = \sum_{n}^N f_n(\alpha_k,t)
\end{align}

which need not necessarily be orthogonal or complete. For example each function might be pulled from a Fourier basis or could be Gaussian shaped pulses, which may lead to different number of parameters per basis function.
\par
To perform a gradient based search over the set of parameters we take the derivative of the objective function w.r.t each parameter

\begin{widetext}

\begin{align}
    \partial_\alpha g &= -\frac{1}{\text{dim(U)}}\bigg[ \Tr(U^\dagger_{target} P_c \partial_\alpha U(\vec{\alpha},T_c) P_c)\Tr(U^\dagger(\vec{\alpha},T_c)U_{target}) \\
    &~~~~~~~~~~~~~~~~~~~+ \Tr(U^\dagger_{target} U(\vec{\alpha},T_c))\Tr(P_c \partial_\alpha U(\vec{\alpha},T_c)^\dagger P_c U_{target}) \bigg]\\
    &= -\frac{2}{\text{dim}(U)} Re\bigg[ \Tr(U^\dagger_{target} P_c \partial_\alpha U(\vec{\alpha},T_c) P_c)\Tr(P_c U^\dagger(\vec{\alpha},T_c) P_c U_{target}) \bigg]
\end{align}

\end{widetext}

where we identify that we must find $\partial_\alpha U(\vec{\alpha},T_c)$. But there doesn't exist an analytic expression for this operator in general. Thus it was proposed to find this quantity via an alternative equation of motion derived from the Schr\"odinger equation. The procedure is to take the Schr\"odinger equation for the unitary operator

\begin{align}
    \partial_t U(t) = \frac{-i}{\hbar} H(t)U(t)
\end{align}

and take its partial derivative w.r.t a parameter $\alpha$

\begin{align}
    \partial_\alpha \partial_t U(t) = \frac{-i}{\hbar} \bigg[ \partial_\alpha H(t)U(t) + H(t) \partial_\alpha U(t) \bigg]
\end{align}

and exchange the order of the second derivative on the left side leading to an equation of motion for the $\partial_\alpha U(\vec{\alpha},T_c)$

\begin{align}
    \partial_t \partial_\alpha U(t) = \frac{-i}{\hbar} \bigg[ \partial_\alpha H(t)U(t) + H(t) \partial_\alpha U(t) \bigg]
\end{align}

which leads to a set of coupled differential equations:

\begin{equation}\label{eq:GOAT_EOMs}
    \frac{\partial}{\partial t}
    \begin{pmatrix}
    U(t)\\
    \partial_\alpha U(t)
    \end{pmatrix} = \frac{-i}{\hbar}
    \begin{pmatrix}
    H(t) & 0\\
    \partial_\alpha H(t) & H(t)
    \end{pmatrix}
    \begin{pmatrix}
    U(t)\\
    \partial_\alpha U(t)
    \end{pmatrix}.
\end{equation}

Thus the GOAT algorithm has two primary steps: (1) solution of the coupled Schr\"odinger equations for every instance of the parameters, and (2) the optimization of the parameters to minimize the objective function $g$. By iterating these steps until a convergence threshold is reached, we are able identify a set of control parameters that are a minimum of the infidelity, Eq.~(\ref{eq:infidelity}). The main advantage to using the GOAT algorithm over other optimal control methods are that the optimized fields are analytic and thus easily interpreted, and potentially easier to calibrate because the number of parameters is independent of the control time, unlike other methods \cite{machnes_tunable_2018}.

\subsection{Implementation}\label{app:implementation}

We implement the GOAT algorithm using the programming language Julia and various open-source packages. Our implementation uses the Julia package DifferentialEquations.jl to numerically solve the coupled GOAT equations of motion, Eq.~(\ref{eq:GOAT_EOMs}) using a order 5/4 Runge-Kutta method with adaptive time stepping \cite{rackauckas2017differentialequations}. For the gradient-based optimization we use a limited-memory Broyden-Fletcher-Goldfarb-Shanno algorithm with a backtracking line-search method, both implemented in the Optim.jl package \cite{mogensen2018optim}. We limit each optimization routine to a maximum of 500 iterations and define a stopping criteria when the infinity-norm of the gradient falls below 1e-5. 

For each of the parameters in Figures \ref{fig:amp_constrained_QSL} and \ref{fig:param_induced_QSL} we run 10 optimization problems each with various initial guesses. The initial guesses for the gradient-based optimization are chosen as the guesses with minimal infidelity observed from an ensemble of 10000 guesses based on a uniform random sampling of each parameter guess range. This large number of initial guesses was chosen to mitigate traps in local minima and because evaluating the infidelity of a control is less computationally expensive than performing optimization using GOAT. For each Gaussian basis function we select an initial guess from the parameter ranges:

\begin{align}
    -0.005\text{ GHz}~\leq ~&a~ \leq~0.003\text{ GHz} \\
    \frac{Tc}{3}\text{ ns} ~\leq ~&\mu~ \leq~ \frac{2T_c}{3}\text{ ns} \\
    1 \text{ ns} ~\leq ~&\sigma~ \leq~ 10 \text{ ns}.
\end{align}

These parameter regimes were selected to bias the optimization a priori towards optimal solutions that are more experimentally feasible, i.e. low amplitude, low frequency controls that start and end smoothly at zero amplitude.

In all of the numerical optimizations performed we constrain the control amplitude to within a range specified by experimental restrictions, shown in Table~\ref{tab:params}. To implement this constraint we construct a saturation function based on a generalized logistic function defined by:

\begin{equation}
    S(\varepsilon(t)) = A - \frac{B-A}{1-Q\exp(-\frac{2g \varepsilon(t)}{B-A})}
\end{equation}

where $A=\min(g_{12}), B=\max(g_{12}), g=4$ are the lower asymptote, upper asymptote, and maximum gradient of the sigmoid, respectively. The additional, variable $Q=-\frac{B}{A}$ is chosen such that $S(0)\approx 0$. Normally, the logistic function is non-constant within a finite domain around $[A,B]$ thus the control field is linearly re-scaled within the  specified amplitude range $[A,B]$ by the ratio of $(B-A)/2$ which gives the magnitude of the input amplitude relative to the desired range. 

This amplitude saturation function acts to bound the field amplitude in a continuous manner and leads to a vanishing gradient component w.r.t the control amplitude when the control amplitude lies outside $[A,B]$, thus implementing a barrier for gradient based optimization. Then, the controls interact with the Hamiltonian as 

\begin{equation}
    H(t) = H_{drift} + \sum_k S_k(\varepsilon_k(t))H_k
\end{equation}

and (for the purposes of the GOAT algorithm) the partial derivative of $H(t)$ w.r.t. a parameter $\alpha$ is given by

\begin{equation}
    \frac{\partial H(t)}{\partial \alpha}\bigg\rvert_t = \frac{\partial S}{\partial \varepsilon}\bigg\rvert_{\varepsilon(t)} \frac{\partial \varepsilon}{\partial \alpha}\bigg\rvert_t H_k.
\end{equation}

\bibliographystyle{unsrt}
\bibliography{Control,Devices,Physics,QuantumSimulation}

\end{document}